\documentclass[lettersize,journal]{IEEEtran}
\usepackage{graphicx}
\usepackage{subfigure}
\usepackage[cmex10]{amsmath}
\usepackage{amssymb}
\usepackage{bm}
\usepackage{algorithm}
\usepackage{algorithmic}
\usepackage{cite}
\usepackage{setspace}
\usepackage{stfloats}
\usepackage{multirow}
\usepackage{mathrsfs}
\usepackage{color}
\usepackage{verbatim}
\usepackage{extpfeil}
\usepackage{hyperref}
\usepackage{caption}
\begin{document}
\makeatletter
\newcommand{\rmnum}[1]{\romannumeral #1}
\newcommand{\Rmnum}[1]{\expandafter \@slowromancap \romannumeral #1@}
\makeatother

\title{Fingerprint Based mmWave Positioning System Aided by Reconfigurable Intelligent Surface}

\author{Tuo Wu, Cunhua Pan, Yijin Pan,  Hong Ren, Maged Elkashlan, and Cheng-Xiang Wang, \emph{Fellow IEEE}

\thanks{(Corresponding author: Cunhua Pan).

T. Wu and M.Elkashlan are with the School of Electronic Engineering and Computer Science at Queen
Mary University of London, London E1 4NS, U.K. (Email:\{tuo.wu, maged.elkashlan\}@qmul.ac.uk).  C. Pan, Y. Pan and H. Ren   are with the National Mobile Communications Research Laboratory, Southeast University, Nanjing 210096, China. C.-X. Wang is with the National Mobile Communications Research Laboratory, Southeast University, Nanjing 210096, China, and also with the Purple Mountain Laboratories, Nanjing, 211111, China. (Email: \{cpan, panyj, hren, chxwang\}@seu.edu.cn).
}

}

\markboth{}
{}

\maketitle

\begin{abstract}
			Reconfigurable intelligent surface (RIS) is a promising technique for millimeter wave (mmWave) positioning systems. In this paper, we consider multiple mobile users (MUs) positioning problem in the multiple-input multiple-output (MIMO) time-division duplex  (TDD) mmWave systems aided by the RIS. We derive the expression for the space-time channel response vector (STCRV) as a novel type of fingerprint. The STCRV consists of the multipath channel characteristics, e.g., time delay and angle of arrival (AOA), which is related to the position of the MU. By using the STCRV as input, we propose a novel residual convolution network regression (RCNR) learning algorithm to output the estimated three-dimensional (3D) position of the MU. Specifically, the RCNR learninng algorithm includes a data processing block to process the input STCRV, a normal convolution block to extract the features of STCRV, four residual convolution blocks to further extract the features and protect the integrity of the features, and a regression block to estimate the 3D position. Extensive simulation results are also presented to demonstrate that the proposed RCNR learning algorithm outperforms the traditional convolution neural network (CNN).
	\end{abstract}
\IEEEpeerreviewmaketitle

\section{Introduction}

Localization-related industries demand high levels of localization accuracy \cite{Pch1}, e.g., mobile user sensing \cite{Zhangw1}.  It is worth pointing out that prevalent global positioning system (GPS) localization accuracy, even in ideal conditions, is approximately 5 meters, which falls short of meeting the stringent requirements of location-sensitive applications. Hence, wireless positioning systems in the millimeter wave (mmWave) band were advocated by some researchers as a way to improve the positioning performance \cite{Wqq3}.  However, due to the sensitivity of mmWave signals to blockages, high-precision positioning is difficult to maintain \cite{Zhi11}.

Reconfigurable intelligent surface (RIS) is an emerging technique for mmWave positioning systems with several advantages \cite{huang2022,Renzo1,Wqq1,Dai5,Wqq2,Zhou1}. First, the RIS can reconstruct a new line-of-sight (LoS) communication link if the direct link was blocked by obstacles \cite{Wqq1}. Second, the RIS provides reliable and high-precision estimation with low energy consumption \cite{Wqq2}. Finally, the RIS saves hardware costs when deploying a wireless positioning reference compared with the access point (AP) \cite{Dai5}. Thus, wireless positioning algorithms aided by the RIS is a promising enabler  for sixth-generation (6G) wireless systems \cite{Zhou1}.

Currently, wireless positioning algorithms aided by the RIS have been studied by some researchers, including two-step positioning algorithms and fingerprint based algorithms \cite{Pyj1,Renzo2}. For two-step positioning algorithms, the channel parameters, e.g., time delay and angle of arrival (AOA), are estimated at the first step. At the second step, the channel parameters can be used to derive the three-dimensional (3D) position of the mobile user (MU) by using the geometry relationship. For instance,  a near-field joint channel estimation and localization algorithm was proposed in \cite{Pyj1}. However, the two-step positioning algorithms depend on the  channel parameters estimation, which require line-of-sight (LoS) measurement. These algorithms may not be suitable for indoor localization as the LoS links may be blocked by obstacles. Without estimating the channel parameters, fingerprint based positioning algorithms directly predict the position by using the fingerprint (e.g., received signal strength information (RSSI)). For example,  \cite{Renzo2} regarded RSSI  as a type of fingerprint to predict the MUs aided by the RIS.  However, RSSI-based fingerprint localization algorithms can be unstable due to the fast fading fluctuation.   Recently, some researchers proposed the channel state information (CSI) as the fingerprint, due to its potential to enhance the positioning accuracy compared with RSSI \cite{3DCNN}.

Against the above background, the main contributions of this paper are summarized as follows:
\begin{itemize}
		\item[1)] For the fingerprint based mmWave positioning system, we propose a new type of fingerprint, space-time channel response vector (STCRV), which consists of multipath channel characteristics. The proposed STCRV fingerprint  is closely related to the position of the MU.
		
		\item[2)] Utilizing the STCRV as the wireless positioning fingerprint, we propose a novel  residual convolution network regression (RCNR) learning algorithm to estimate the 3D positions of the MUs. Specifically,  STCRV is processed by a data processing block at first. Then, a normal convolution block is then used  to extract the features of the output from the data processing block. Consequently,  four residual convolution blocks are utilized to further extract the features and protect the integrity. Finally,  the 3D position is estimated through a regression block.
		
		\item[3)] Simulation results are provided to evaluate the performance of the proposed RCNR learning algorithm. The proposed algorithm outperforms the CNN in terms of root mean square error (RMSE).
		
	\end{itemize}

\section{System Model and Problem Formulation} \label{System Model}
We consider an RIS-aided 3D massive  multiple-input multiple-output (MIMO) time-division duplex (TDD) mmWave positioning system, where the MUs send pilot signals to the AP to locate the positions of the MUs aided by an RIS. In addition, we assume that the direct channels between the AP and the MUs are blocked by some obstacles, such as thick walls..

\begin{figure}[!ht]
	\centering
	{\includegraphics[width=2.6in]{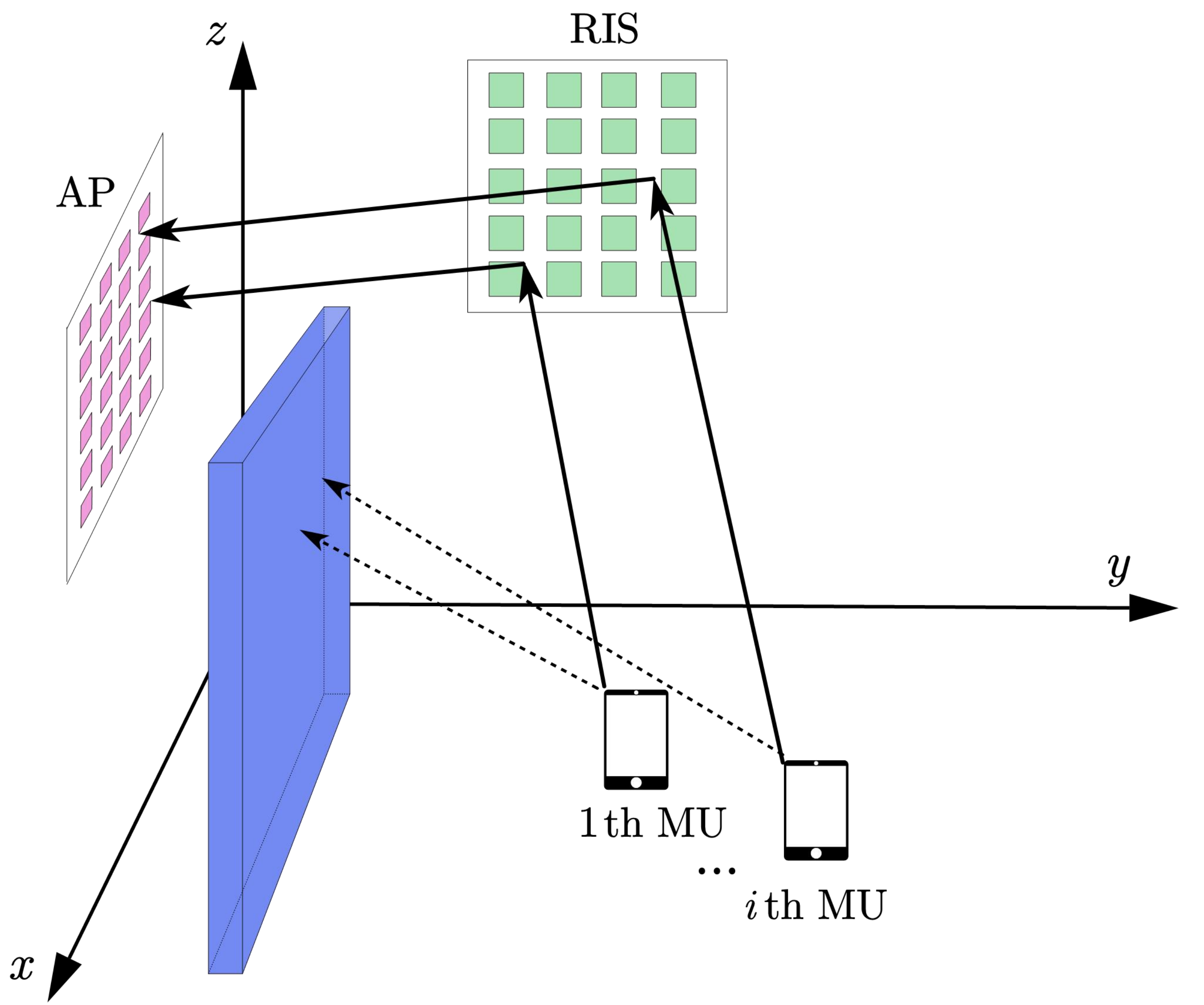}}
	\caption{\normalsize RIS-aided positioning system model.}\label{model}
\end{figure}

 As we can see from Fig. \ref{model}, the AP is placed on the left side of the wall with the center located at ${\bf p}=[x_p,y_p,z_p]^T$. Moreover, the AP is assumed to be equipped with a uniform planar array (UPA) with  $M_{x,z}=M_x\times M_z$ antennas, where $M_x$ and $M_z$ denote the numbers of antennas along the x-axis and the  z-axis, respectively.  Additionally, the RIS is assumed to be placed at the front side of the wall with the center located at ${\bf s}=[x_s,y_s,z_s]^T$. The UPA-based RIS has $N_{y,z}=N_y\times N_z$ reflecting elements along the y-axis and the z-axis, respectively. Furthermore, there are $N_u$ MUs , each of which is equipped with a single antenna. The  MUs are located at ${\bf u}_i=[x_i,y_i,z_i]^T$, $i=1,2,\cdots,N_u$.

It is assumed  that the number of propagation paths between the $i$th MU and the RIS is $N_p$, the AOA of the $p$th path from the $i$th MU to the RIS can be decomposed into the elevation angle $0\leq\theta_{p,i}\leq\pi$ in the vertical direction, and the azimuth angle $0\leq\phi_{p,i}\leq\pi$ in the horizontal direction.  As a result, the array response vector at the RIS can be expressed as \cite{wang10pervasive}
\begin{align}\label{1}
{\bf a}_{R_a}(\theta_{p,i},\phi_{p,i})={\bf a}_{R_a}^{(e)}(\theta_{p,i})\otimes{\bf a}_{R_a}^{(a)}(\theta_{p,i},\phi_{p,i}),
\end{align}
where $\otimes$ denotes the  the Kronecker product. Moreover, we have
\begin{align}\label{2}
{\bf a}_{R_a}^{(e)}(\theta_{p,i})=[1,e^{\frac{-j2\pi d_r\cos\theta_{p,i}}{\lambda_c}},...,e^{\frac{-j2\pi(N_{y}-1) d_r\cos\theta_{p,i}}{\lambda_c}}]^T,
\end{align}
and
\begin{align}\label{3}
{\bf a}_{R_a}^{(a)}(\theta_{p,i},\phi_{p,i})=&[1,e^{\frac{-j2\pi d_r\sin\theta_{p,i}\cos\phi_{p,i}}{\lambda_c}},...,\nonumber\\
&e^{\frac{-j2\pi(N_{z}-1) d_r\sin\theta_{p,i}\cos\phi_{p,i}}{\lambda_c}}]^T,
\end{align}
where $d_r$ and $\lambda_c$ denote the distance between the adjacent elements of the RIS and   the carrier wavelength, respectively. Then, the channel from the $i$th MU to the RIS, denoted as ${\bf g}_i$,  can be modeled as
\begin{align}\label{4}
{\bf g}_i = \sum^{N_p}_{p=1}\alpha_{p,i}{\bf a}_{R_a}(\theta_{p,i},\phi_{p,i}),
\end{align}
where $\alpha_{p,i}$ denotes the complex channel gain of the $p$th path.
\begin{figure*}
\centering
	{\includegraphics[width=7in]{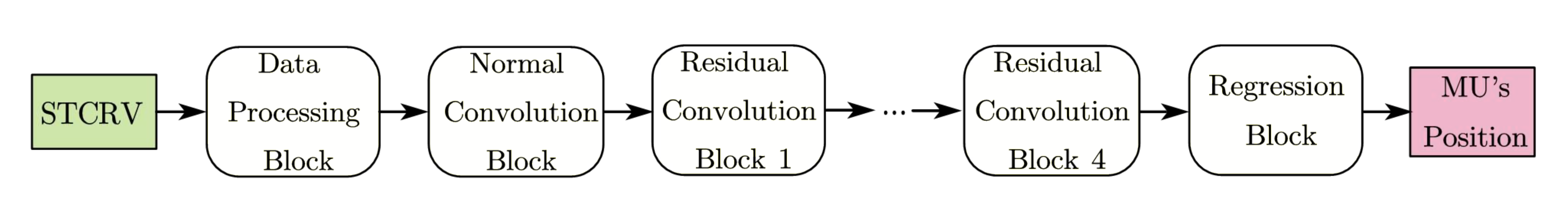}}
	\caption{\normalsize The structure of the residual convolution network regression (RCNR) learning algorithm.}\label{algorithm}
\end{figure*}

Similarly, it is assumed that the number of propagation paths between the AP and the RIS is $N_j$. The  angle of departure (AOD) of the $j$th path from the RIS to the AP can be decomposed into the elevation angle $0\leq\theta_{j}\leq\pi$ in the vertical direction and the azimuth angle $0\leq\phi_{j}\leq\pi$ in the horizontal direction. Hence, the array response vector ${\bf a}_{R_d}(\theta_{j},\phi_{j})$ can be written as
\begin{align}\label{5}
{\bf a}_{R_d}(\theta_{j},\phi_{j})={\bf a}_{R_d}^{(e)}(\theta_{j})\otimes{\bf a}_{R_d}^{(a)}(\theta_{j},\phi_{j}),
\end{align}
with
\begin{align}\label{6}
{\bf a}_{R_d}^{(e)}(\theta_{j})=[1,e^{\frac{-j2\pi d_r\cos\theta_{j}}{\lambda_c}},...,e^{\frac{-j2\pi(N_{y}-1) d_r\cos\theta_{j}}{\lambda_c}}]^T,
\end{align}
and
\begin{align}\label{7}
{\bf a}_{R_d}^{(a)}(\theta_{j},\phi_{j})=&[1,e^{\frac{-j2\pi d_r\sin\theta_{j}\cos\phi_{j}}{\lambda_c}},...,\nonumber\\
&e^{\frac{-j2\pi(N_{z}-1) d_r\sin\theta_{j}\cos\phi_{j}}{\lambda_c}}]^T.
\end{align}

For the RIS-AP link,  the AOA of the $j$th path can be decomposed into the elevation angle $0\leq\psi_{j}\leq\pi$ in the vertical direction and the azimuth angle $0\leq\omega_{j}\leq\pi$ in the horizontal direction. Therefore, the array response vector ${\bf a}_{B}(\psi_{j},\omega_{j})$ can be written as
\begin{align}\label{8}
{\bf a}_{B}(\psi_{j},\omega_{j})={\bf a}_{B}^{(e)}(\psi_{j})\otimes{\bf a}_{B}^{(a)}(\psi_{j},\omega_{j}),
\end{align}
with
\begin{align}\label{9}
{\bf a}_{B}^{(e)}(\psi_{j})=[1,e^{\frac{-j2\pi d_b\cos\psi_{j}}{\lambda_c}},...,e^{\frac{-j2\pi(M_{x}-1) d_b\cos\psi_{j}}{\lambda_c}}]^T,
\end{align}
and
\begin{align}\label{10}
{\bf a}_{B}^{(a)}(\psi_{j},\omega_{j})=&[1,e^{\frac{-j2\pi d_b\sin\psi_{j}\cos\omega_{j}}{\lambda_c}},...,\nonumber\\
&e^{\frac{-j2\pi(M_{z}-1) d_b\sin\psi_{j}\cos\omega_{j}}{\lambda_c}}]^T,
\end{align}
where $d_b$ denotes the distance of the antennas of the AP.

 By using the array response vector ${\bf a}_{R_d}(\theta_{j},\phi_{j})$ in  \eqref{5} and ${\bf a}_{B}(\psi_{j},\omega_{j})$ in \eqref{8}, the channel matrix of the AP with the RIS can be formulated as
\begin{align}\label{11}
{\bf H}=\sum_{j=1}^{N_j}\beta_{j}{\bf a}_{R_d}(\theta_{j},\phi_{j}){\bf a}^{H}_{B}(\psi_{j},\omega_{j}),
\end{align}
where $\beta_{j}$ denotes the channel gain of the $j$th path.

Denote ${\bf\Psi}_t\in\mathbb{C}^{N_{y,z}\times N_{y,z}}$ as the phase shift matrix of the RIS in time slot $t$. It is assumed that the MUs transmit pilot sequences of length $\tau$ via the RIS to the AP. During the uplink transmission of the $i$th MU, in time slot $t$,  $1\leq t \leq\tau$, the received signal from the $i$th MU at the AP can be written as
\begin{align}\label{12}
{\bf y}_i(t) = {\bf H}{\bm \Psi}_t{\bf g}_i\sqrt{p}s_i(t)+{\bf n}_i(t),
\end{align}
where $s_i(t)$  denotes the pilot signal from the $i$th MU, ${\bf n}_i(t)\in\mathbb{C}^{M_{x,z}\times1}$ $ \thicksim$ $\mathcal{CN}(0,\delta^2{\bf I})$ represents additive white Gaussian noise (AWGN) with power $\delta^2$ at the AP. Here,  $p$ denotes the transmit power of the $i$th MU.  According to the expression of the received signal from the $i$th MU, we can define the space-time channel response vector (STCRV) of the $i$th MU as
\begin{align}\label{13}
{\bf h}_i = {\bf H}{\bm \Psi}_t{\bf g}_i,
\end{align}
which consists of the channel parameters (e.g., channel gain and AOA/AOD).  The STCRVs are unique for different positions, hence the STCRVs can be regarded as a new type of  CSI  fingerprint of the MU.  According to the definition of directly positioning algorithms \cite{3DCNN}, the STCRVs can be used to estimate the positions of the MUs. Therefore, by denoting the estimated position of the $i$th MU as $\hat{\bf u}_i=[\hat{x}_i,\hat{y}_i,\hat{z}_i]^T$, we have
 \begin{align}\label{14}
\hat{\bf u}_i=f({\bf h}_i),
\end{align}
where $f(\cdot)$ denotes the complex non-linear function between the STCRV and the estimated position of the $i$th MU.
Hence, the regression problem of estimating the positions of the MUs can be formulated as
\begin{align}\label{15}
\min_{\hat{\bf u}_i\atop i=1,\cdots,N_u}\quad \frac{1}{N_u}\sum_{i=1}^{N_u}(\hat{\bf u}_i-{\bf u}_i)^2.
\end{align}

\section{Residual Convolution Network Regression Learning  Algorithm}
According to the regression problem in \eqref{15}, the closed-form expression of the 3D position of the MU is not available by using traditional optimization methods.  Hence, a regression learning algorithm to predict the positions of MUs is developed in this section.  Deep learning based method, popular in image recognition of computer science, can be applied to the STCRV fingerprint positioning since the  STCRV can be seen as an image.  Therefore, we propose a novel residual convolution network regression (RCNR) learning algorithm to represent the  function $f(\cdot)$ in \eqref{14} and solve the  regression problem \eqref{15}.
\subsection{Regression-oriented Positioning}
In computer science, convolution neural network (CNN) is often used for image classification with the last layer being activated by a softmax function \cite{CNN}. As an alternative to the classification function, CNN can also be viewed as a regression function if the softmax layer is replaced by a fully connected layer containing an activation function. As an advanced network of CNN, residual convolution network (RCN) can also be used as regression function by substituting a fully connected layer to the last softmax layer.
\begin{figure}[!t]
\centering \subfigure[Data processing (DP) block.]{  \includegraphics[width=0.50\linewidth]{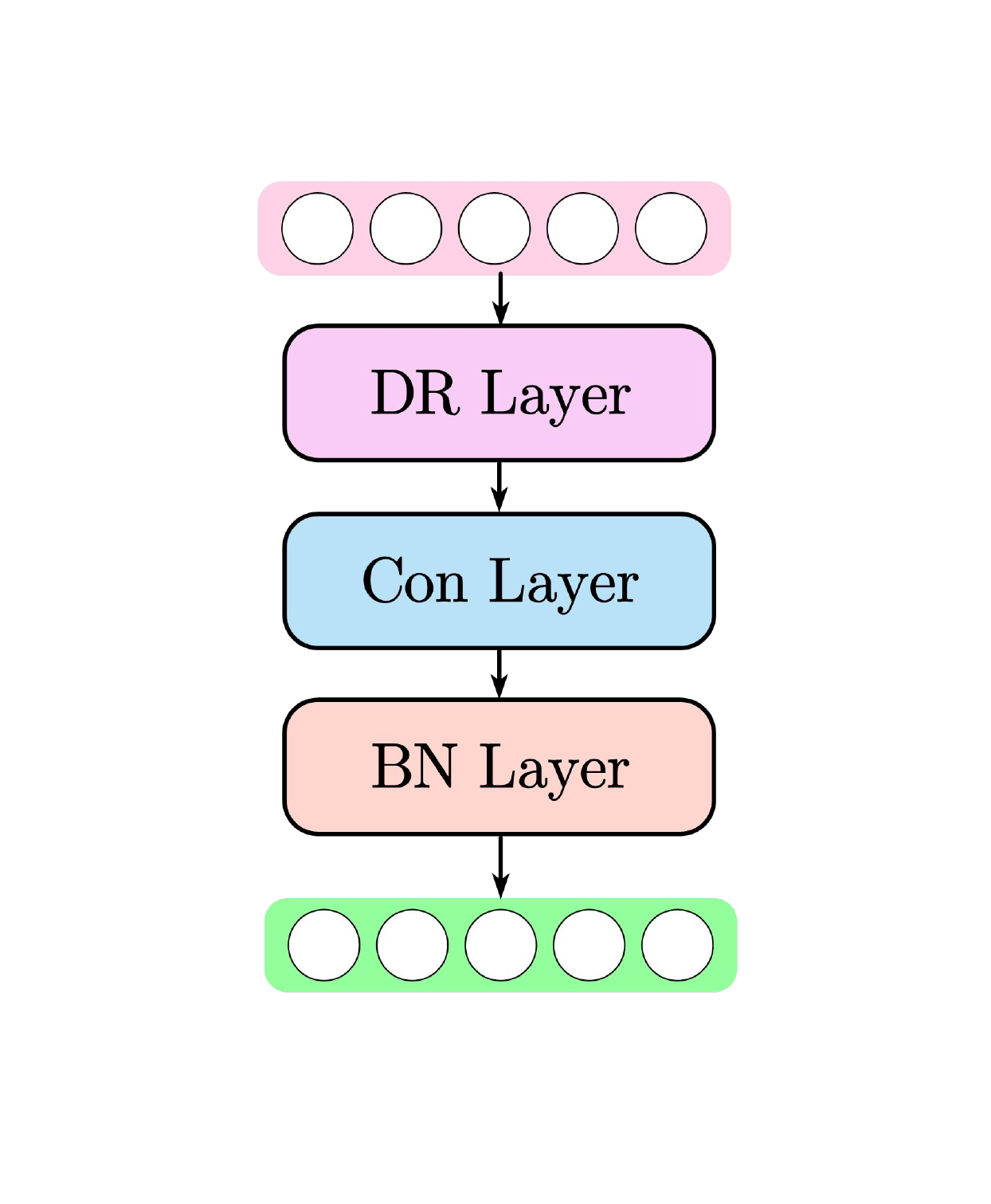}\label{DP}
}\subfigure[Regression block.]{  \includegraphics[width=0.50\linewidth]{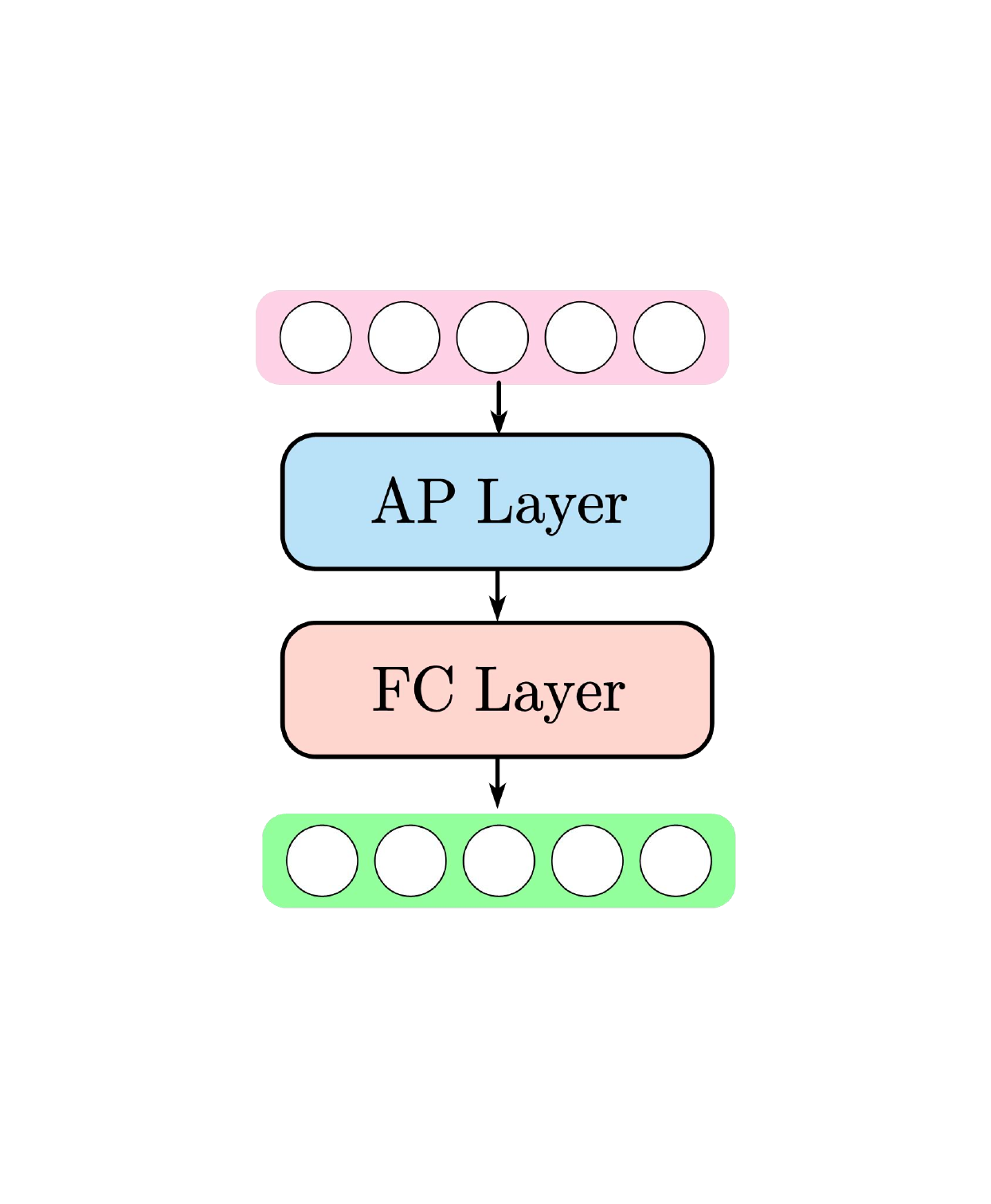}\label{Output}
}\caption{\normalsize Two blocks used for the proposed algorithm.}
\label{convolution}
\end{figure}

\subsection{The structure of the RCNR learning algorithm}
Fig. \ref{algorithm} shows the details of the proposed algorithm. As shown in Fig. \ref{algorithm}, the proposed algorithm consists of a data processing block, a normal convolution block, four residual convolution blocks, and a regression block. The descriptions of these blocks will be introduced as follows.

\subsubsection{Data Processing Block}
The data processing (DP) block is designed to process the input STCRV. As we can see from Fig. 3-(a), the DP block includes two layers: a data decomposing (DD) layer, and a data reshape (DR) layer.

First, since the STCRV is a complex vector, we design the DD layer to decompose ${\bf h}_i$ into two parts, the real value vector ${\bf h}^{(r)}_i$ and the imaginary valure vector ${\bf h}^{(i)}_i$, which can be expressed as
  \begin{align}\label{17}
{\bf h}_i={\bf h}^{(r)}_i+j{\bf h}^{(i)}_i.
\end{align}

Then,  according to the expression of the STCRV in \eqref{13},  the STCRV consists of the features of the horizontal angle domain and the vertical angle domain of the AP. To further extract these features, we design the DR layer to reshape the two vectors ${\bf h}^{(r)}_i\in \mathbb{C}^{M_{x,z}\times 1}$ and ${\bf h}^{(i)}_i\in \mathbb{C}^{M_{x,z}\times 1}$ as two space-time channel response matrices (STCRM), which are denoted as ${\bf H}^{(r)}_i\in \mathbb{C}^{M_{x}\times M_{z}}$ and ${\bf H}^{(i)}_i\in \mathbb{C}^{M_{x}\times M_{z}}$, respectively. To be specific, we reshape these two vectors by arranging the elements of vector into $M_{x}$  rows and $M_{z}$ columns.
\subsubsection{Normal Convolution Block}

\begin{figure}[!t]
\centering \subfigure[Normal convolution (NC) block.]{  \includegraphics[width=0.50\linewidth]{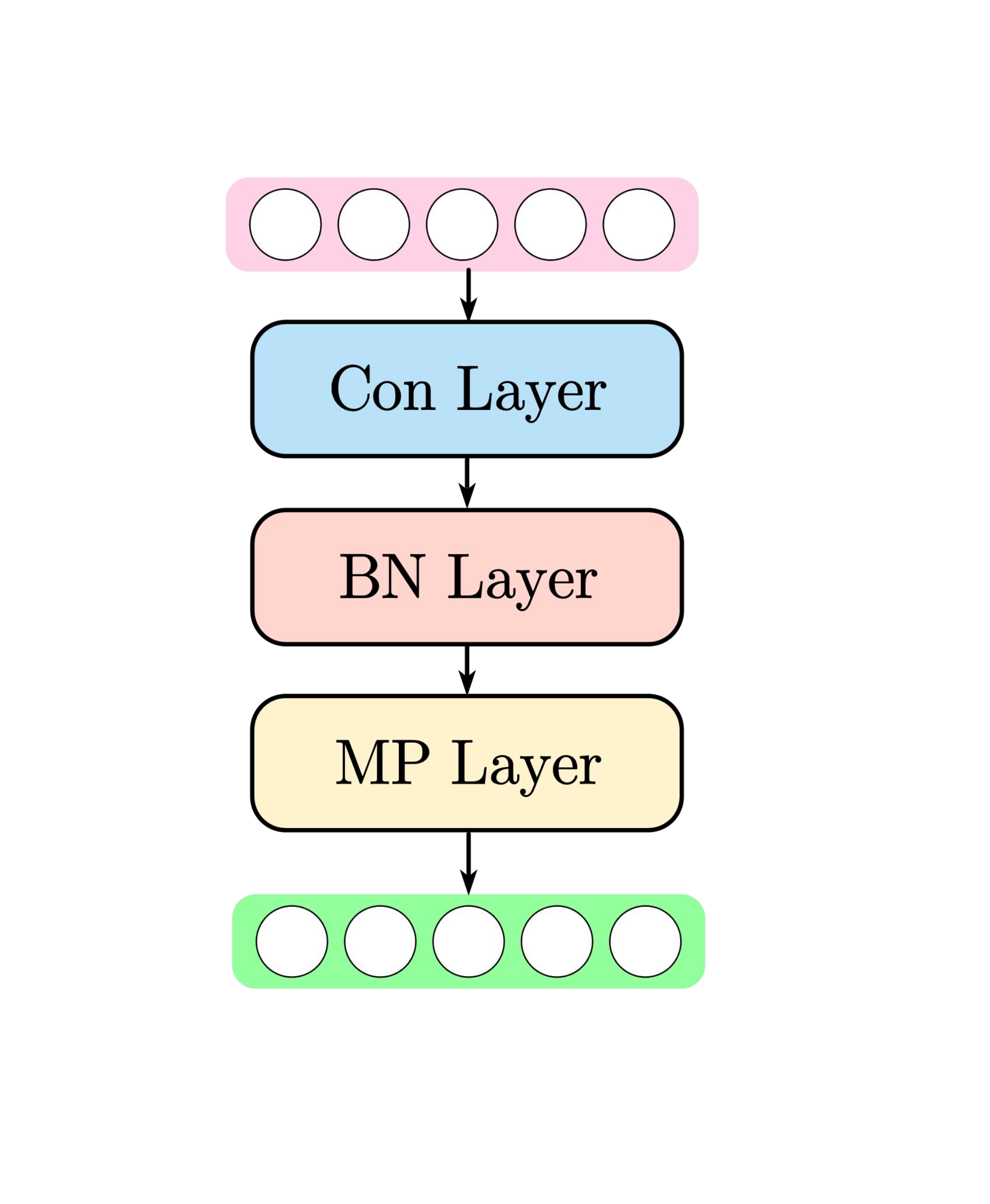}\label{NC}
}\subfigure[Residual convolution (RC) block.]{  \includegraphics[width=0.50\linewidth]{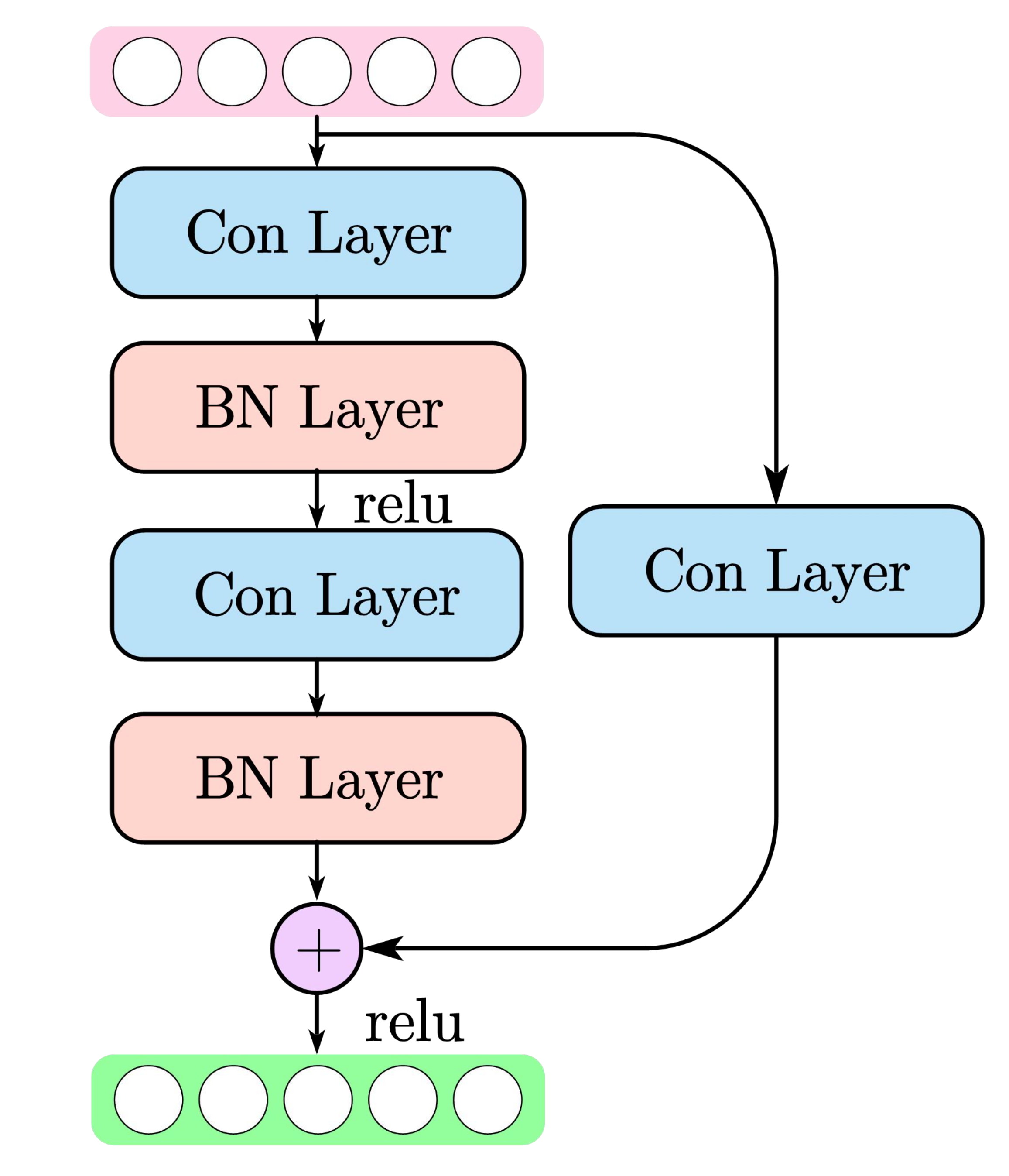}\label{residual}
}\caption{\normalsize Two kinds of convolution block used for the proposed algorithm.}
\label{convolution}
\end{figure}

The normal convolution (NC) block is designed to extract the features of the STCRM.  As we can see from Fig. 4-(a), the NC block consists of three layers: a convolution (Con) layer, a batch normalization (BN) layer, and a max pooling (MP) layer.

Due to the large number of antennas at the AP, the STCRM has a large dimension. Therefore, if the STCRM is input directly into a deep neural network (DNN) consisting of fully connected layers,  a large number of weight parameters of the DNN should be trained. To improve the efficiency of the training neural network, the Con layer is proposed \cite{CNN}. The Con layer has multiple filters sliding over it for a given input STCRM so that the features of STCRM can be extracted \cite{CNN}.  As a result, the number of weight parameters to be trained can be reduced.

Moreover,  when training DNN, the input for each layer changes as the weight parameters of the previous layers change,  leading to  a reduction in the convergence rate.  Hence, the BN layer is designed to normalize the input data so that the convergence rate can be improved.

Furthermore, the MP layer is designed to reduce the complexity of further layers, which is similar to reducing the resolution in the field of computer science.

\subsubsection{Residual Convolution Block}
In the field of computer science, increasing the number of the Con layers will allow the neural network to extract more features \cite{CNN}. However, according to the experiment in \cite{Resnet}, when the neural network reaches a certain depth, the problem of gradient explosion and gradient disappearance will appear, which leads to a worse optimization effect and lower accuracy of the proposed neural network. Hence, to improve the estimation accuracy,  the residual convolution network is proposed to enable the deeper neural network to train well and obtain a better optimization effect \cite{Resnet}.

Inspired by the residual convolution network of computer science, the residual convolution (RC) block is designed to further extract the features of STCRV and protect the integrity of features in our proposed RCNR learning algorithm. As shown in Fig. 4-(b).  The RC block includes three Con layers and two BN layers.

 As we can see from Fig. 4-(b),  the RC block starts with two Con layers. Each Con layer is followed by a BN layer and a ReLU activation function. Then we skip these 2 convolutional operations through the cross-layer datapath and add the input directly before the final ReLU activation function.   As a result,  the integrity of the features is protected and the degradation of the neural network can be solved.

\subsubsection{Regression Block}
The regression block is designed to output the estimated positions of the MUs. As shown in Fig. 3-(b), the regression block includes an average pooling (AP) layer and a fully connected (FC) layer.

The AP layer is used to reshape the output of the RC blocks for the final FC layer by taking the average of each feature from the RC block \cite{CNN}. Adding the AP layer between the RC block and the FC layer avoids the large number of weight parameters introduced by the FC layer.  As a result, the AP layer reduces overfitting while improving the convergence rate.

 The FC layer is used to combine the features from the NC blocks and output the estimated positions of the MUs. The FC layer multiplies the input by a parametric weight matrix and then adds a bias vector.  By using ${\bm \Omega}$ to denote the output before the FC layer, the estimated position after the FC layer can be written as
\begin{align}\label{16}
\hat{\bf u}_i={\bf W}\textrm{vec}\{{\bm \Omega}\}+{\bf b},
\end{align}
where ${\bf W}$ and ${\bf b}$ are the parametric weight matrix and bias vector respectively that can be learned together with the training of the RCNR learning algorithm.

\section{Simulation Results} \label{result}

In this section, simulation results are provided to evaluate the performance of the proposed RCNR algorithm. The software Wireless Insite \cite{wi} is used to simulate the mmWave positioning system aided by the RIS.   For the AP, the number of antennas is set to $M_{x,z}=M_x\times M_z=255\times255$. Besides, the center of the AP is located at ${\bf p}=(-10, -5, 2.5 $ m$)$, and the distance between the antennas $d_b$ is set to 0.2 m. For the RIS, the number of the elements is set to $N_{y,z}=N_y \times N_z=255 \times 255$. Moreover, the center of the RIS is located at ${\bf s}=(-5.10 , -1.43, 2 $ m$)$ and the distance of the elements $d_r$ is set to $0.2$ m. For the MUs, we assume that the MUs are uniformly distributed in the grid of $9.6$ m in length and $5.8$ m in width. Furthermore, we have set up three grids in total, and the heights of the grids are $1.4$ m, $1.5$ m, and $1.6$ m, respectively. In addition, the distance of the MUs is $0.2$ m and the transmit power of the MUs is $10$ dBm.  The Ray tracing propagation model is selected to simulate the communication system from the MUs via the RIS to the AP. The phase shifts of the elements at the RIS are set to a unity matrix.

\begin{figure}[!ht]
	\centering
	{\includegraphics[width=2.6in]{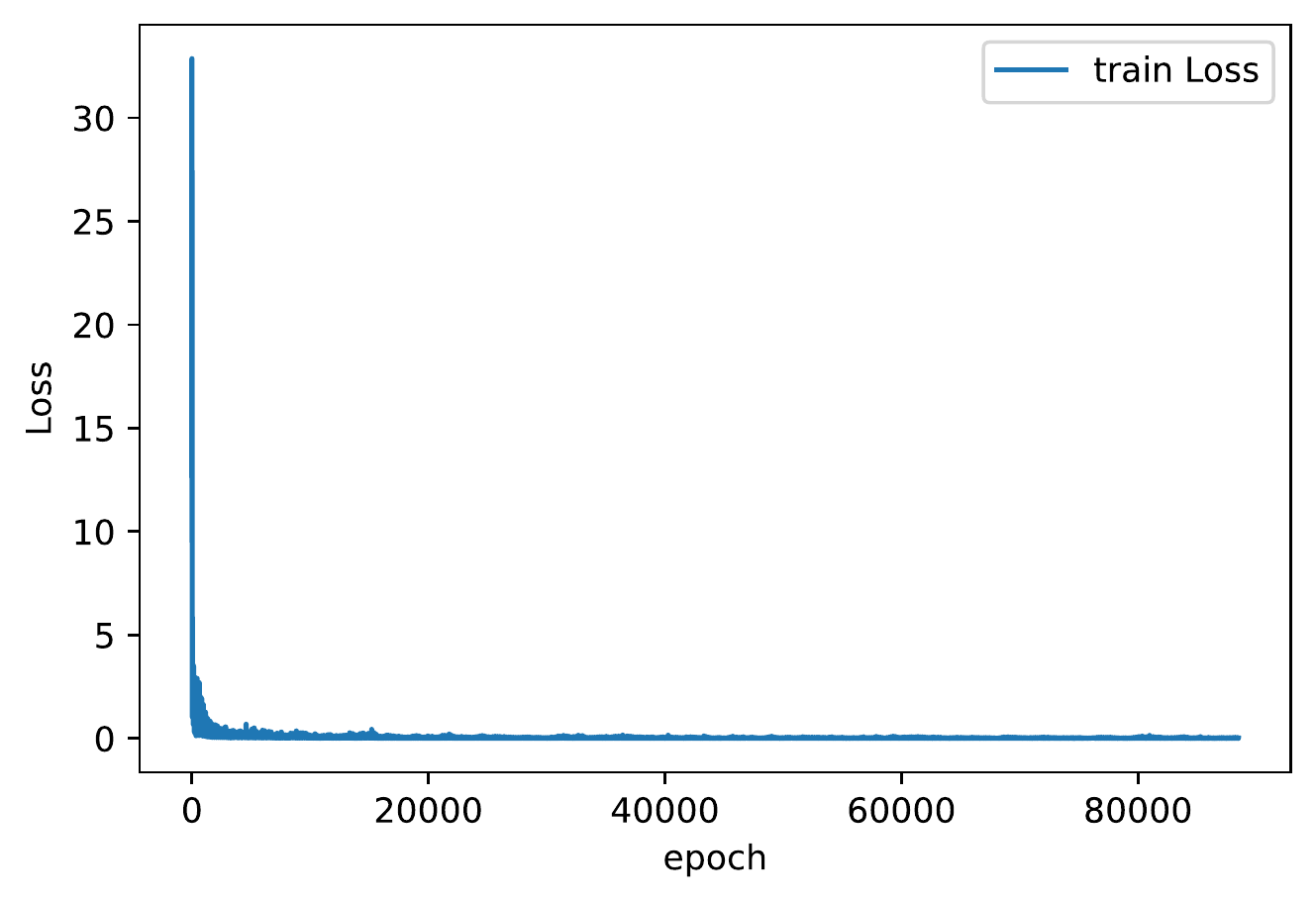}}
	\caption{\normalsize Train Loss of the RCNR learning algorithm.}\label{train_Loss}
\end{figure}

\begin{figure}[!ht]
	\centering
	{\includegraphics[width=2.6in]{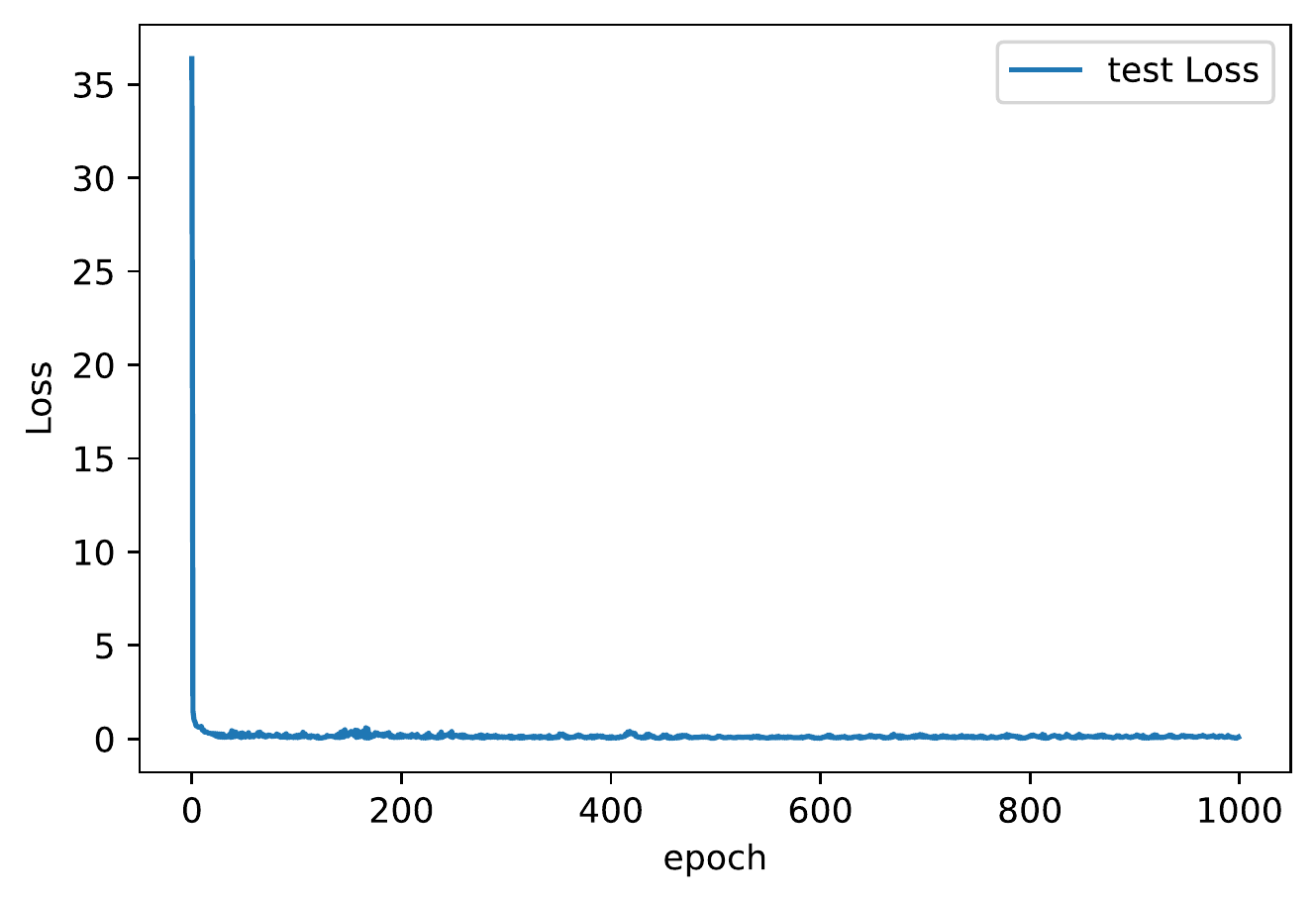}}
	\caption{\normalsize Test Loss of the RCNR learning algorithm.}\label{test_loss}
\end{figure}

In Fig. \ref{train_Loss} and Fig. \ref{test_loss}, the convergence behavior of the proposed RCNR algorithm is demonstrated. Both Fig. \ref{train_Loss} and Fig. \ref{test_loss} have one curve, named `train Loss' and `test Loss', respectively. These curves denote the train loss and the test loss of the proposed RCNR learning algorithm. The loss can be obtained by
  \begin{align}\label{18}
Loss = \frac{1}{N}\sum_{({\bf x},{\bf y})\in D}({\bf y}-\hat{\bf y})^2,
\end{align}
where $D$ denotes the dataset, $({\bf x},{\bf y})$ denotes the sample of the dataset,  ${\bf x}$  represents the input STCRV and ${\bf y}$ denotes the 3D coordinate of the MU. $\hat{\bf y}$ denotes the prediction of the 3D coordinate of the MU. $N$ denotes the number of samples of the dataset.

It can be observed from Fig. \ref{train_Loss} and Fig. \ref{test_loss} that both train loss and test loss decrease when the number of epochs increases, which means that the proposed RCNR algorithm is learning to estimate the positions of the MUs.

\begin{figure}[!ht]
	\centering
	{\includegraphics[width=2.8in]{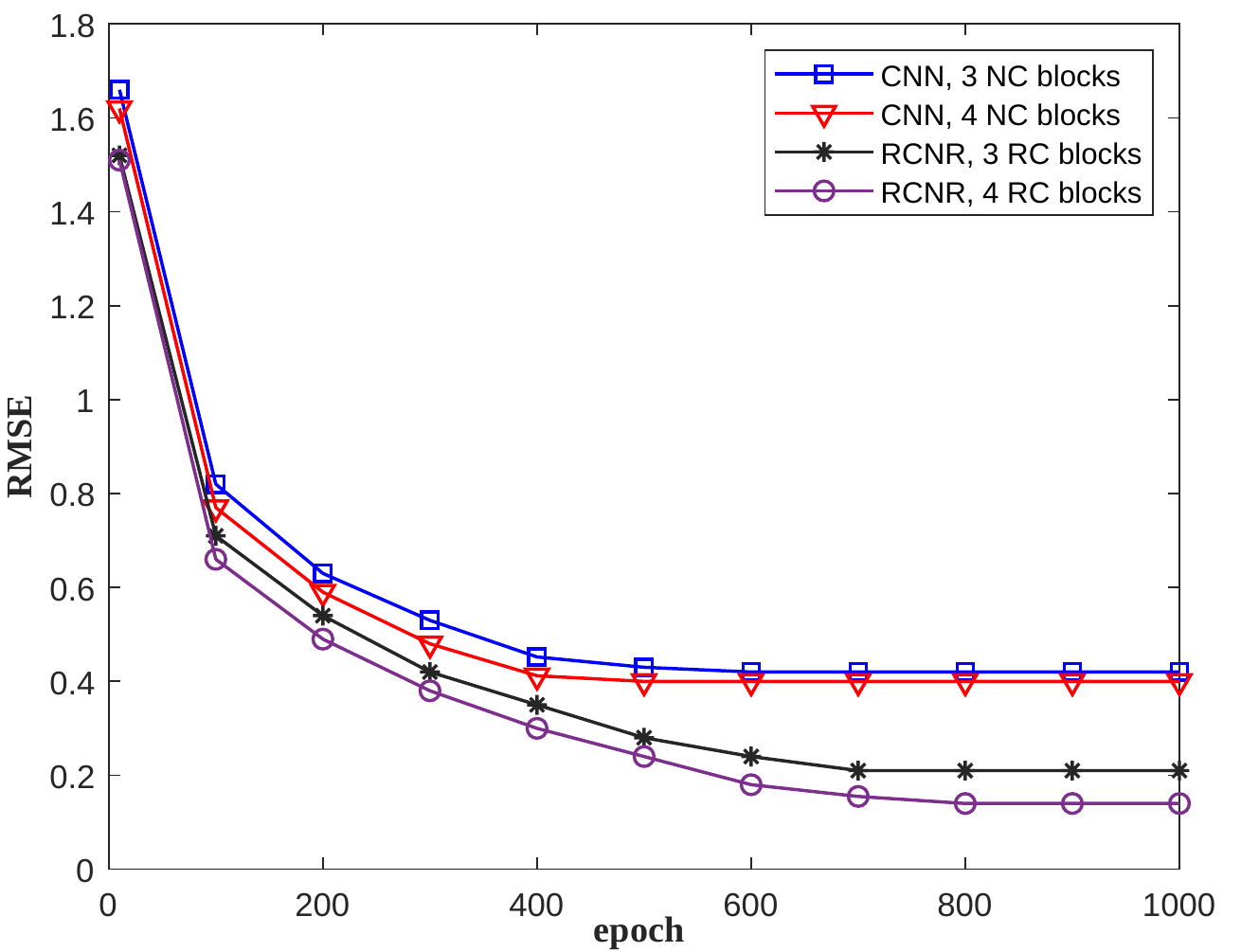}}
	\caption{\normalsize Test RMSE of the RCNR learning algorithm.}\label{test_RMSE}
\end{figure}
To evaluate the performance of the proposed RCNR algorithm, we present the RMSE comparison of the RCNR algorithm with 3 RC  blocks, the RCNR algorithm with 4 RC blocks, the CNN algorithm with 3 NC  blocks, and the CNN algorithm with 4 NC blocks, denoted as `CNN, 3 NC blocks', `CNN, 4 NC blocks', `RCNR, 3 RC blocks', and `RCNR, 4 RC blocks'. As shown in Fig. \ref{test_RMSE}, with the increase in the number of convolution blocks, the RMSE becomes lower, which means that the proposed algorithm with more RC blocks can achieve higher accuracy. Furthermore, the proposed RCNR algorithm outperforms the CNN algorithm. With more convolution blocks, our proposed RCNR algorithm performs better than CNN.

\section{Conclusion}\label{Con}
In this paper, we studied the MU positioning  problem in MIMO TDD mmWave systems aided by the RIS. We derived the expression for STCRV at the AP as a new type of fingerprint. In addition, by using the STCRV fingerprint as input, we proposed a novel RCNR algorithm to predict the 3D position of the MU. Extensive simulation results were presented to demonstrate the superiority of the proposed RCNR algorithm over the CNN algorithm.

\bibliographystyle{IEEEtran}
\bibliography{myre}
\end{document}